\documentclass{article}
\usepackage{spconf,amsmath,graphicx}
\usepackage{enumitem}
\usepackage{subfig}
\usepackage{amsfonts}
\usepackage{amsmath}
\usepackage{amssymb}
\usepackage{hyperref}
\usepackage{url}

\DeclareMathOperator*{\argmin}{arg\,min}
\title{COMPOSITIONAL EMBEDDING MODELS FOR SPEAKER IDENTIFICATION AND DIARIZATION WITH SIMULTANEOUS SPEECH FROM 2+ SPEAKERS}
\name{Zeqian Li and Jacob Whitehill\thanks{This material is based on work supported by the National Science Foundation  under grants \#1822768 and \#2019805.}}
\address{Worcester Polytechnic Institute (WPI), Massachusetts, USA}
\begin{document}
%\ninept
%
\maketitle
\begin{abstract}
We propose a new method for 
speaker diarization that can handle overlapping speech with 2+ people. Our method is based on \emph{compositional
embeddings} \cite{li2020compositional}: Like standard speaker embedding methods such as x-vector \cite{snyder2018x}, compositional embedding models contain a function $f$ that separates speech from different speakers. In addition, they include a composition function $g$ to compute set-union operations in the embedding space so as to infer the \emph{set} of speakers within the input audio. In an experiment on multi-person speaker  identification using synthesized LibriSpeech data, the proposed method outperforms traditional embedding methods that are only trained to separate single speakers (not speaker sets). In a speaker diarization experiment on the AMI Headset Mix corpus, we achieve state-of-the-art accuracy (DER=22.93\%), slightly better than the previous best result (23.82\% from \cite{bullock2020overlap}).
\end{abstract}
\section{Introduction}
\label{sec:intro}

Real-world speech often includes moments with simultaneous speech from multiple speakers, and an important task in automated speech analysis is to identify not just the single speaker, but rather the entire \emph{set} of speakers, who is speaking at each moment in time. This gives rise to the computational problems of (a) multi-person speaker identification and (b) speaker diarization with simultaneous speech.
For \emph{single-speaker} identification and diarization -- i.e., where at most one person is speaking at a time -- the state-of-the-art approach is based on deep speaker embedding models such as x-vector \cite{snyder2018x}.  Embedding methods seek to map examples (e.g., an MFCC feature vector from an audio frame) into a metric space such that examples from the same class (speaker) are close together and examples from different classes are far apart. However, they are fundamentally limited because they assume that only a single speaker is speaking at any time.%, and hence it is unclear how they can be used for multi-person identification and diarization.

In this paper, we harness a recently proposed method for one-shot learning called \emph{compositional embeddings} \cite{li2020compositional,alfassy2019laso} to infer the \emph{set} of people who is speaking within an input audio. Compositional embeddings extend single-speaker embeddings through  a composition function that is trained to estimate the location in the embedding space of where the \emph{union} of two (or more) speakers is located. 
By composing the embedded one-shot examples (i.e., a sample of each person speaking in isolation) and comparing the result to the embedding of the test audio, the set of speakers can be inferred.

{\bf Contributions}: (1) We propose novel methods for multi-person speaker identification, and for speaker diarization with simultaneous speech, based on compositional embeddings. (2) We improve upon the original compositional embedding model of Li et al.~\cite{li2020compositional} by modifying its normalization method. (3) We show that the proposed diarization method achieves state-of-the-art accuracy, slightly outperforming the previous best result \cite{bullock2020overlap} on the AMI Headset Mix corpus.

\section{Related Work}
{\bf Single-speaker  identification and diarization}:
The state-of-the-art for these problems  is based on embedding models, which can either be probabilistic subspace models such as i-vector \cite{dehak2011language}, or deep neural network-based models such as x-vector \cite{snyder2018x} and d-vector \cite{variani2014deep}. The latter can be trained to minimize a cross-entropy loss over a fixed set of training classes \cite{snyder2018x}, or to minimize an embedding loss (e.g., triplet  \cite{weinberger2009distance}, generalized end-to-end  \cite{wan2018generalized}) in a one-shot learning setting.

{\bf Compositional embeddings}: 
The computer vision community has recently explored compositional embeddings for object recognition problems in which each image can contain multiple objects. Li et al.~\cite{li2020compositional} and Alfassy et al.~\cite{alfassy2019laso} both proposed models containing separate \emph{embedding} and \emph{set union} functions, similar to our approach in this paper. In contrast to \cite{li2020compositional}, our paper uses a different normalization method that we found generally gives higher accuracy (see Section \ref{sec:norm}).% In contrast to \cite{alfassy2019laso}, who use a strong supervision signal with a fixed set of classes during training, our work uses a one-shot learning paradigm with episodic training.

{\bf Multi-person speaker diarization}: The most common approach to  speaker diarization with simultaneous speech is to use an overlapping speech detector; for those segments that contain overlap, the set of speakers can be estimated \cite{bullock2020overlap,yella2014overlapping,zelenak2011detection,zelenak2012simultaneous, fujita2019end}. For the latter step, one approach is  to select the top $k$  closest speakers in the embedding space. In contrast, compositional embedding models can jointly predict both the number of speakers and their identities. 
%Moreover, it is not  it generalizes to any number of simultaneous speakers. %While some methods employ multiple channels (e.g., audio + video)  \cite{minotto2015multimodal}, our approach requires only a single audio channel.

\section{Compositional Embeddings for Speaker Identification \& Diarization}

\subsection{Single-Speaker Identification and Diarization}
For the setting when at most one person is speaking at a time, the state-of-the-art approach to speaker identification and diarization is based on speaker embedding models. The set of speakers at test time is generally not known during training. For speaker identification, one enrollment example  of each speaker in isolation is given to the model.
The goal is to infer, in any given input audio, the single person who is speaking. For speaker diarization, there are no explicit enrollment examples; however, the process of computing distances for clustering is similar to comparing test examples with enrollment examples in identification.

Let $\mathcal{S}$ be the set of all possible speakers who may be speaking in a given audio.
Let $x \in  \mathbb{R}^n$ be the input audio %(or a  feature representation such as MFCC)
, and let $y(x) \in \mathcal{S}$ be the single speaker contained in $x$.  %Zeqian: shouldn't here be "one speaker" instead of "set of speakers"?
We assume that, for each $s\in\mathcal{S}$, we have one enrollment example $\bar{x}_s$ of person $s$ speaking in isolation. Let $d$ be a distance metric such as Euclidean distance, negative cosine similarity, etc.
We wish to train an embedding function $f: \mathbb{R}^n \rightarrow \mathbb{R}^m$ (where $m$ is the dimension of the embedding space) with the property that $d(f(x_a), f(x_b))$ is small if $y(x_a)=y(x_b)$ and is large if $y(x_a)\ne y(x_b)$.
Commonly used loss functions   include the triplet loss \cite{weinberger2009distance} and generalized end-to-end loss \cite{wan2018generalized}.

{\bf Inference}: Given a trained $f$, we can determine the speaker $s$ in a test example $x$ by comparing $f(x)$ with the embedded enrollment examples. First, define $\bar{e}_s = f(\bar{x}_s)$. Then 
$\hat{y}(x) = \argmin_{s\in \mathcal{S}} d(f(x), \bar{e}_s)$.
Although we focus on one-shot learning, this formulation can  be extended to the case with multiple enrollment examples per speaker $s$ (few-shot learning) by computing the  distance of $f(x)$ to the centroid of the embedded enrollments for each speaker $s$. %For simplicity, our paper focuses on the one-shot case.

\subsection{Multi-Speaker Identification and Diarization}
\label{sec:format}
\begin{figure}
    \centering
    \includegraphics[width=0.8\columnwidth]{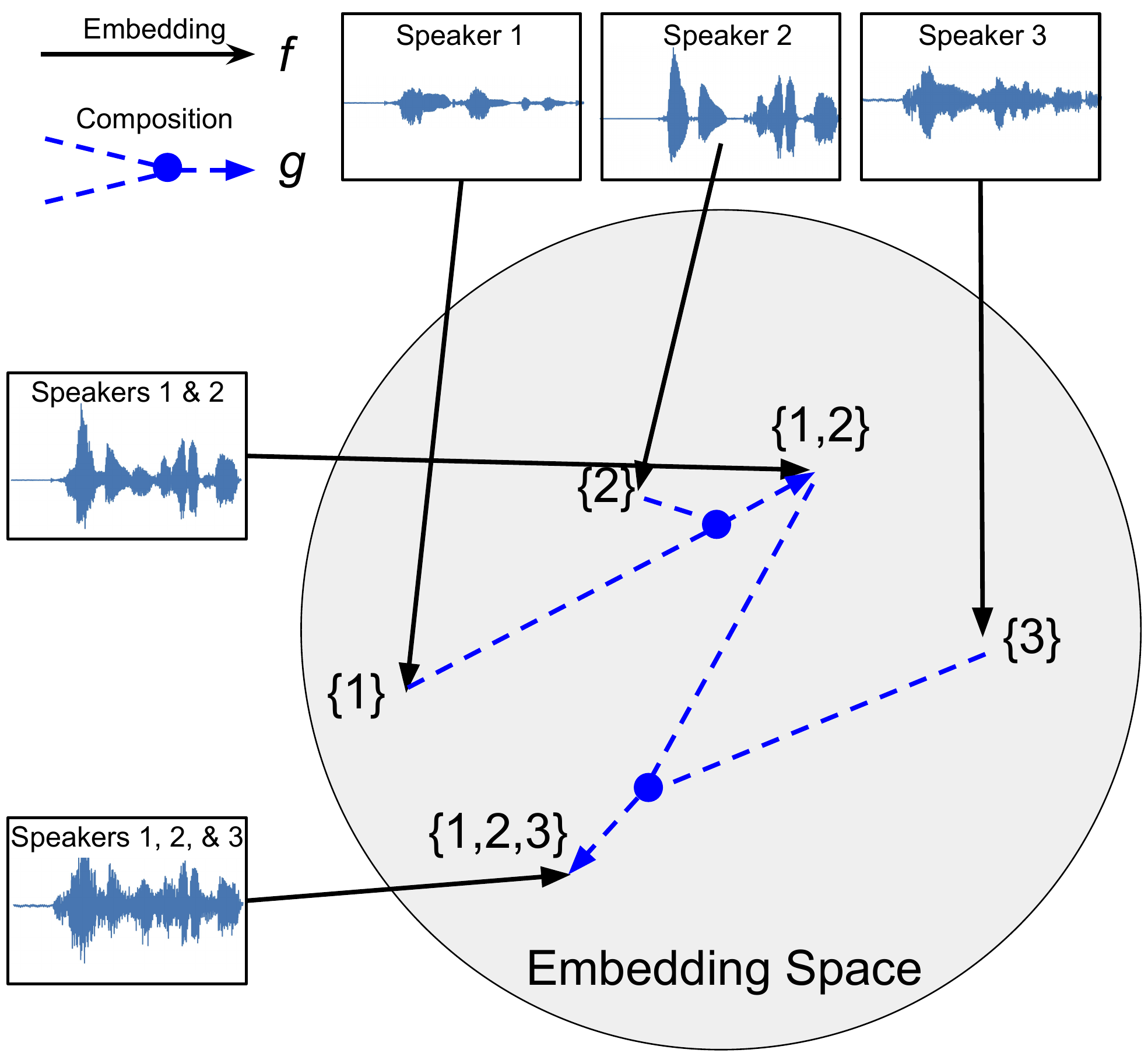}
    \caption{Compositional embedding model with embedding function $f$ and composition function $g$: given one example each of speakers 1, 2, and 3, the location in the embedding space of any set of speakers can be estimated with $g$.}
    \label{fig:compemb}
\end{figure}
Here we describe the compositional embedding approach that we use for speaker identification and diarization with simultaneous speech from 2+ speakers. Our model is based on work by \cite{li2020compositional}, who originally proposed the method for multi-object image recognition. Function $f$ is an embedding model that takes a data sample as input and outputs an embedding represents the input data sample. Function $g$ takes 2 embeddings as input and outputs an embedding of the composition of the 2 data samples represented by the 2 input embeddings. The outputs of both $f$ and $g$ are used in loss function thus that $f$ and $g$ are optimized jointly. 
In contrast to the single-speaker setting, here the assumption is that, 
at any moment in time, a subset $\mathcal{T} \subseteq \mathcal{S}$ may be speaking simultaneously. Hence, we redefine $y$ to output the \emph{set} of people speaking, i.e., $y(x) \subseteq \mathcal{S}$. As with single-speaker embedding models, we train $f$ so that $d(f(x_a), f(x_b))$ is small if $y(x_a)=y(x_b)$ and is large if $y(x_a)\ne y(x_b)$. However, in contrast to the single-person setting, both individual speakers and any combination thereof must all reside at distinct locations within the embedding space. In addition, we also  train a \emph{composition function} $g: \mathbb{R}^m \times \mathbb{R}^m \rightarrow \mathbb{R}^m$ that consumes two embedding vectors (each corresponding to a particular subset of speakers) and outputs the vector in the same embedding space corresponding to the \emph{union} of the speaker sets in its two inputs. In other words,
for any $x_a,x_b,x_{ab}$ such that $y(x_{ab}) = y(x_a) \cup y(x_b)$, we want 
$g(f(x_a), f(x_b))\approx f(x_{ab})$
(see Figure \ref{fig:compemb}). Note that $g$ can be called recursively many times, e.g., 
$g(f(x_c), g(f(x_a), f(x_b)))\approx x_{abc}$, where $x_{abc}= x_a \cup x_b \cup x_c$. We train $f$ and $g$ \emph{jointly} so as to encourage $f$ to produce embeddings that can be composed with each other using $g$.

{\bf Inference}: Given trained $f$ and $g$, we can estimate the set $\mathcal{T}$ of speakers contained in any input $x$: We first compute the \emph{pseudo-enrollments} of each \emph{subset} $\mathcal{T} \subseteq{S}$. %(or at least those subsets which are deemed feasible, e.g., with at most $k$ speakers, for desired $k$).
Pseudo-enrollment $\bar{e}_\mathcal{T}$ denotes the location in the embedding space representing all the speakers in $\mathcal{T}$ speaking simultaneously. Suppose
$\mathcal{T} = \{ s_1,\ldots,s_p: s_i \in \mathcal{S} \}$ contains $p>1$ speakers. Then $\bar{e}_\mathcal{T}$ can be computed recursively as
\[
\bar{e}_\mathcal{T} = g\left(f(\bar{x}_{s_p}), \bar{e}_{\mathcal{T}\setminus \{s_p\}}\right)
\]
%In Figure \ref{fig:compemb}, the location marked ``$\{1,2,3\}$''  is the pseudo-enrollment  for speakers 1, 2, and 3 speaking simultaneously.

Finally, if $x$ is the the test audio, then we compute its embedding $f(x)$; then, given  the enrollment and pseudo-enrollment examples, we find
$\hat{y}(x) =
\argmin_{\mathcal{T}\subseteq \mathcal{S}} d(f(x), \bar{e}_\mathcal{T})$.
In other words, we find the speaker set $\mathcal{T}$ whose enrollment (or pseudo-enrollment) example $\bar{x}_\mathcal{T}$ is closest to the embedding of the audio input $x$. In the worst case, we must search over $2^{|\mathcal{S}|}$ possible subsets. However, in practical diarization settings, the set of possible speakers, as well as maximum the number of overlapping speakers (e.g., $|\mathcal{T}| \leq 3$), is typically small enough so that this iteration is very tractable.

%{\bf Remarks}: The order in which we recursively apply $g$ to the elements of $\mathcal{T}$ can  affect the result. The reason is that, even if $g$ is defined to be symmetric, it is not associative. In practice, we apply the recursion according to an arbitrary but fixed ordering on $\mathcal{S}$.  Note that the pseudo-enrollment embeddings need to be computed only once (whenever $\mathcal{S}$ changes).

\subsection{Normalizing the Embedding Vectors}
\label{sec:norm}
The original  model  by Li et al.~\cite{li2020compositional} employed an $L_2$ normalization layer as the last layer of the neural networks of $f$ and $g$. This forces the output to lie on a sphere, which is a common embedding technique \cite{schroff2015facenet} and helps to prevent the embedding vectors from exploding in magnitude. However, given that $g$ may be called multiple times when computing the pseudo-enrollment examples, normalizing the output can slow training  due to dampened gradients, and  reduce the representational capacity of the  model.
In our work, we revise the  model to omit the normalization layer from the last layers of $f$ and $g$. Instead we    normalize  just before inference; hence, to infer the speaker set, we find $\hat{y}(x) =
\argmin_{\mathcal{T}\subseteq \mathcal{S}} d(z(f(x)),  z(\bar{e}_\mathcal{T}))$
where $z(\cdot)$ normalizes its input to length 1.
%We found that this strategy usually gave  higher accuracy.

\section{Experiment 1: Multi-Person Speaker Identification (LibriSpeech)}
\label{sec:g_librispeech}
Here we assess the accuracy of  compositional embeddings to identify the set of speakers in fixed-length (2sec) audios.

{\bf Dataset}:
For training and testing, we synthesize audio  with simultaneous speech from 1-3 speakers by combining clips from  LibriSpeech  \cite{panayotov2015librispeech}.
The LibriSpeech training sets (train-clean-100, train-clean-360, and train-other-500), validation set (dev-clean), and test sets (test-clean and test-other) contain a total of 2338, 40, and 73 speakers, respectively. We generate 2-second audios by randomly picking utterances from 1-3 speakers, adding their individual waveforms component-wise, and then normalizing for   volume\footnote{Listen to examples at \url{https://drive.google.com/drive/folders/1L4RIhu49NHmL_hV4_x0wMOZRRe-clSHi}.}. From each synthesized clip, we extract MFCC features (32 coefficients, 0.025s window size, 0.01s step size) to yield the input $x$.

{\bf Models}:
We compared the following different models:
\begin{enumerate}[leftmargin=0.5cm,itemsep=0mm]
    \item {\bf CmpEmL2}: Compositional embeddings where the last layers of $f$ and $g$  perform $L_2$ normalization (like in \cite{li2020compositional}).
    For $f$, we used a 2-layer LSTM with 256 hidden units following by a 256-to-32 dense layer, and then normalized to length 1. Composition network $g$ was defined as $g(x_a, x_b) = z(W_1 x_a + W_1 x_b + W_2 (x_a \odot x_b))$, where $W_1,W_2$ are learned weight matrices.
    
    \item {\bf CmpEm}: Same as \#1 above, except we use the normalization method described in Section \ref{sec:norm}.
    
    \item {\bf SingleEm}: Single-speaker embedding models have no inherent mechanism to represent simultaneous speech. Nonetheless, we can   estimate the speaker set in a given audio $x$ by finding  $\argmin_{\mathcal{T}\subseteq \mathcal{S}} d\left(f(x), \frac{1}{|\mathcal{T}|}\sum_{s\in\mathcal{T}}\bar{e}_s\right)$.
    For example, to test whether $x$ contains speaker $a$, speaker $b$, or speakers $a\&b$, we compare $f(x)$ to $\bar{e}_a,\bar{e}_b,\bar{e}_{ab}$ and output the set that minimizes the distance. Alternatively, if $|\mathcal{T}|=k$ is already known, we can pick the  $k$ speakers whose enrollments $\bar{e}_s$ are closest to $f(x)$; this is the method used  in \cite{bullock2020overlap} during overlapping speech.
    
    \item {\bf Guess}: Baseline for guessing. 
    %Since there are $20+5$ possible speaker sets (for $|\mathcal{S}|=5$ and $|\mathcal{T}|\leq 3$) and an equal probability of selecting each speaker set $\mathcal{T}$, the baseline accuracy for guessing the speaker set  is $100\%/25=4\%$ for an exact match (top-1) and $3\times 4\%=12\%$ (top-3).
\end{enumerate}

{\bf Training}:
{\em CmpEm, CmpEmL2}:
To train   $f$ and  $g$, we use episodic training, where each  episode 
has 5 different speakers (i.e., $|\mathcal{S}|=5$) with at most 3 people speaking at once (i.e., $|\mathcal{T}| \leq 3$). Since ${5 \choose 3} + {5 \choose 2} = 20$, there are 20 pseudo-enrollment + 5 enrollment examples for each episode. We generate 100,000 episodes for training set, 1,000 episodes for the validation set and 10,000 episodes for test set.
Training is performed using Adam ($\textrm{lr}=.0003$) to maximize the validation accuracy. We use triplet loss  with $\epsilon=0.1$.  Functions $f$ and $g$ are trained jointly, with gradients  backpropagated through $g$ to $f$.
{\em SingleEm}:  We trained a single-speaker embedding  $f$ with the same architecture as in CmpEm; it achieved a test accuracy on identifying the single speaker (from a set of 5) of $95.3\%$.

{\bf Evaluation}:
The test set contains examples from speakers not seen during training; they are uniformly distributed across the 25 speaker sets (5 singletons, 10 2-sets, and 10 3-sets).
We assessed test accuracy   in 4 ways:
\begin{enumerate}[leftmargin=0.5cm,itemsep=0mm]
\item Accuracy (\%-correct), over all examples, of inferring $\mathcal{T}$.
\item Accuracy, over examples for which $|\mathcal{T}|=k$ ($k\in\{1,2,3\}$), of inferring $\mathcal{T}$, when $|\mathcal{T}|$ is \emph{not given} to the model and must thus be inferred. %This result can reveal whether a model is more accurate on examples with fewer vs.~more speakers.
\item Accuracy, over all examples, in determining just the \emph{number} of classes in the set, i.e., $|\mathcal{T}|$. In particular, if $|\mathcal{T}|>1$, then overlapping speech is detected.
\item Accuracy, over examples for which $|\mathcal{T}|=k$, when $|\mathcal{T}|$ is \emph{given} to the model by an oracle.
\end{enumerate}

\begin{table}
\begin{center}
\begin{tabular}[t]{c||c|c|c|c}
\multicolumn{5}{c}{\bf Experiment 1: Multi-Person Spkr ID on LibriSpeech} \\\hline \hline
{\em \#spkrs} & \multicolumn{4}{c}{\em Model} \\\hline
 &   CmpEmL2 & CmpEm & SingleEm & Guess \\
\hline\hline
\multicolumn{5}{c}{\bf Speaker Set Identification When $|\mathcal{T}|$ is Unknown} \\\hline
$\leq 3$      & 72.1 & {\bf 74.9} & 37.5 &  4.0 \\ 
 \hline
1   & 92.6 & {\bf 93.5} & 77.6 & 4.0 \\    \hline
2   & {\bf 75.9} & 75.6 & 34.4 & 4.0 \\  
    \hline
3   & 57.9 & {\bf 64.8} & 20.6 & 4.0 \\  
  \hline \hline
\multicolumn{5}{c}{\bf Speaker Set Size Estimation} \\\hline
$\leq 3$       & 92.5 & {\bf 94.3} & 56.5 & 36.0
\\\hline \hline
\multicolumn{5}{c}{\bf Speaker Set Identification When $|\mathcal{T}|$ is Given} \\ \hline
1 & - & 94.1 & {\bf 96.6} & 20.0\\ \hline
2 & - & {\bf 81.4} & 54.1 & 10.0\\ \hline
3 & - & {\bf 68.0} & 33.2 & 10.0
\end{tabular}
\end{center}
\caption{{\bf Experiment 1 (LibriSpeech)}:
Mean accuracy (\% correct) in inferring the speaker set $\mathcal{T}$ in each audio, as well as the number of speakers $|\mathcal{T}|$.}
\label{tbl:LibriResult}
\end{table}
{\bf Results}:
{\em $|\mathcal{T}|$ is unknown}:
Results are given in Table \ref{tbl:LibriResult}.
For inferring $\mathcal{T}$, since there are 25 different possible speaker sets, the baseline for guessing is $4\%$.
While SingleEm is much better than chance, it is worse than both compositional embedding methods ($37.5\%$ vs.~$74.9\%$ and $72.1\%$). This underlines the fact that single-speaker embeddings have no inherent ability to identify the size or identity of \emph{sets} of speakers. Between the two compositional embedding methods, CmpEm generally does better. For inferring $|\mathcal{T}|$, the guess rate is higher since there are only three possibilities (1,2,3), but the trend among  methods is the same (Guess $<$ SingleEm $<$ CmpEm).

{\em $|\mathcal{T}|$ is given}: The SingleEm slightly outperforms CmpEm when $|\mathcal{T}|=1$, which is expected since it focuses exclusively on modeling individual speakers. However, it performs much worse than CmpEm on simultaneous speech ($|\mathcal{T}|\geq 2$).

\section{Experiment 2:  Speaker Diarization (AMI)}
Here we apply compositional embeddings to speaker diarization with simultaneous speech from multiple speakers.

{\bf Dataset}:
We use  AMI Headset Mix  \cite{carletta2007unleashing},  one of the largest  diarization datasets. In the test set, 81\% of the  speech is non-overlapping, and 19\% is overlapping \cite{bullock2020overlap}.

{\bf Models}:
In all models, the embedding function $f$  has the same architecture (but different learned weights): From each audio input $x$ (2sec long, step size of 0.5sec), SincNet features \cite{ravanelli2018speaker} are extracted and  passed to an x-vector network \cite{snyder2018x}, followed by a 512-dim embedding layer. Also, all models begin with a Voice Activity Detector (VAD)  and then a Speaker Change (SCD) Detector to find speaker turns. We compare the following approaches, some of which operate on speaker turns and some on 1sec audio segments: 
\begin{enumerate}[leftmargin=0.5cm,itemsep=0mm]
    \item 
    {\bf SingleEm (turn)}:
    Long ($\geq 3.3$sec) turns are diarized by computing the mean embedding of each and then clustering the mean embeddings using affinity propagation \cite{yin2018neural}; this yields the centroids for all the speakers.  Each short ($<3.3$sec) turn is diarized by averaging its embeddings and then assigning it   to the  speaker centroid with highest cosine similarity. Note that this model always predicts a single speaker. This approach is from \cite{yin2018neural}.
    
    \item {\bf SingleEm (segment) with overlap detector}: After SCD, run the overlapping speech detector  to find segments with multiple speakers. Instead of assigning whole turns to speakers, divide them into 1sec segments. Assign each non-overlapping segment to closest single speaker, and assign each overlapping segment to the 2 closest speaker centroids. This approach is very similar to and achieves equal accuracy as \cite{bullock2020overlap}.
    %Our approach is different
    
    \item {\bf CmpEm (segment)}: Similar to \#1, long turns are diarized by clustering their average embeddings into speakers. The centroid embedding for each speaker becomes its enrollment $\bar{e}_s$, and enrollments are composed using $g$ to compute pseudo-enrollments. Then, all turns are divided into 1sec segments, and each segment is assigned to a speaker set by maximizing cosine similarity of its embedding to $\mathcal{T} \in \{ \bar{e}_\mathcal{T}:  |\mathcal{T}| \leq 2 \}$. Hence, this method can infer the set of speakers without a dedicated overlap detector.
    
    \item {\bf CmpEm (segment) with overlap detector}: Same as \#3, but also run the overlap detector from \cite{bullock2020overlap}. Split each turn into 1sec segments. Assign each overlapping segment by maximizing cosine similarity of its mean embedding from  $\{ \bar{e}_\mathcal{T}: |\mathcal{T}|=2\}$; assign each non-overlapping segment by maximizing cosine similarity to 1-speaker enrollments.
\end{enumerate}

{\bf Training}\footnote{Code for all experiments:
\url{https://drive.google.com/drive/folders/1zml_dSQL9RdZ-ggju5ba3M5zQJttSWxo}}: We jointly train the $f\&g$ in the two CmpEm models above using VoxCeleb1 \cite{nagrani2017voxceleb} and VoxCeleb2 \cite{chung2018voxceleb2} training sets augmented with MUSAN \cite{musan2015} background noise and music. The training procedure is the same as in Experiment 1. The $f$  in the SingleEm models is from pyannote-audio \cite{Bredin2020} and pretrained using the same datasets. The VAD model, SCD, and overlap detector are all from pyannote-audio and pretrained using AMI training set.
% ZEQIAN -- please cite the correct sources for these datasets.
% PLEASE PROVIDE MORE TRAINING DETAILS TOO. Done.

{\bf Evaluation}:
We assess  Diarization Error Rate (DER) -- the fraction of the audio  attributed to the wrong speaker(s).

\begin{table}
    \centering
    \begin{tabular}{l|c}
    \multicolumn{2}{c}{\bf Experiment 2: Spkr Diarization on AMI Headset Mix}\\\hline\hline
    Method & DER\% \\ \hline
    SingleEm (turn) \cite{yin2018neural} & 30.12\\
    SingleEm (segment) with overlap detector \cite{bullock2020overlap} & 23.82\\
    CmpEm (segment) & 25.97 \\
    CmpEm (segment) with overlap detector & {\bf 22.93}
    \end{tabular}
    \caption{Diarization Error Rate (DER\%) of different speaker diarization methods on the AMI Headset Mix test set.}
    \label{tab:my_label}
\end{table}
{\bf Results}:
The CmpEm (segment) with the overlap detector gave the lowest diarization error ($22.93\%$) and slightly outperformed the SingleEm (segment) with overlap detector method ($23.82\%$, which matches the number reported in \cite{bullock2020overlap}).
Without the use of an overlap detector, the CmpEm (segment) can jointly determine the size and identities in the speaker set for each segment: the DER is substantially better than the SingleEm approach without an overlap detector, though not as good as with the dedicated detector. Overall, the results suggest that  compositional embedding models can  increase accuracy for speaker diarization with overlapping speech.

\section{Conclusions}
We have shown promising results of  compositional embeddings for  speaker identification and diarization with simultaneous speech. Future work can examine how to cluster speaker embeddings and diarize them  in a single pass, rather than the multi-stage approach we used in CmpEm (segment).

\bibliographystyle{IEEEbib}
\bibliography{strings,refs}

\begin{thebibliography}{10}

\bibitem{li2020compositional}
Zeqian Li, Mike Mozer, and Jacob Whitehill,
\newblock ``Compositional embeddings for multi-label one-shot learning,''
\newblock {\em arXiv preprint arXiv:2002.04193}, 2020.

\bibitem{snyder2018x}
David Snyder, Daniel Garcia-Romero, Gregory Sell, Daniel Povey, and Sanjeev
  Khudanpur,
\newblock ``X-vectors: Robust dnn embeddings for speaker recognition,''
\newblock in {\em 2018 IEEE International Conference on Acoustics, Speech and
  Signal Processing (ICASSP)}. IEEE, 2018, pp. 5329--5333.

\bibitem{bullock2020overlap}
Latan{\'e} Bullock, Herv{\'e} Bredin, and Leibny~Paola Garcia-Perera,
\newblock ``Overlap-aware diarization: Resegmentation using neural end-to-end
  overlapped speech detection,''
\newblock in {\em ICASSP 2020-2020 IEEE International Conference on Acoustics,
  Speech and Signal Processing (ICASSP)}. IEEE, 2020, pp. 7114--7118.

\bibitem{alfassy2019laso}
Amit Alfassy, Leonid Karlinsky, Amit Aides, Joseph Shtok, Sivan Harary, Rogerio
  Feris, Raja Giryes, and Alex~M Bronstein,
\newblock ``Laso: Label-set operations networks for multi-label few-shot
  learning,''
\newblock in {\em Proceedings of the IEEE Conference on Computer Vision and
  Pattern Recognition}, 2019, pp. 6548--6557.

\bibitem{dehak2011language}
Najim Dehak, Pedro~A Torres-Carrasquillo, Douglas Reynolds, and Reda Dehak,
\newblock ``Language recognition via i-vectors and dimensionality reduction,''
\newblock in {\em Twelfth annual conference of the international speech
  communication association}, 2011.

\bibitem{variani2014deep}
Ehsan Variani, Xin Lei, Erik McDermott, Ignacio~Lopez Moreno, and Javier
  Gonzalez-Dominguez,
\newblock ``Deep neural networks for small footprint text-dependent speaker
  verification,''
\newblock in {\em 2014 IEEE International Conference on Acoustics, Speech and
  Signal Processing (ICASSP)}. IEEE, 2014, pp. 4052--4056.

\bibitem{weinberger2009distance}
Kilian~Q Weinberger and Lawrence~K Saul,
\newblock ``Distance metric learning for large margin nearest neighbor
  classification.,''
\newblock {\em Journal of Machine Learning Research}, vol. 10, no. 2, 2009.

\bibitem{wan2018generalized}
Li~Wan, Quan Wang, Alan Papir, and Ignacio~Lopez Moreno,
\newblock ``Generalized end-to-end loss for speaker verification,''
\newblock in {\em 2018 IEEE International Conference on Acoustics, Speech and
  Signal Processing (ICASSP)}. IEEE, 2018, pp. 4879--4883.

\bibitem{yella2014overlapping}
Sree~Harsha Yella and Herv{\'e} Bourlard,
\newblock ``Overlapping speech detection using long-term conversational
  features for speaker diarization in meeting room conversations,''
\newblock {\em IEEE/ACM Transactions on Audio, Speech, and Language
  Processing}, vol. 22, no. 12, pp. 1688--1700, 2014.

\bibitem{zelenak2011detection}
Martin Zelen{\'a}k and Javier Hernando,
\newblock ``The detection of overlapping speech with prosodic features for
  speaker diarization,''
\newblock in {\em Twelfth Annual Conference of the International Speech
  Communication Association}, 2011.

\bibitem{zelenak2012simultaneous}
Martin Zelenak, Carlos Segura, Jordi Luque, and Javier Hernando,
\newblock ``Simultaneous speech detection with spatial features for speaker
  diarization,''
\newblock {\em IEEE Transactions on Audio, Speech, and Language Processing},
  vol. 20, no. 2, pp. 436--446, 2012.

\bibitem{fujita2019end}
Yusuke Fujita, Naoyuki Kanda, Shota Horiguchi, Kenji Nagamatsu, and Shinji
  Watanabe,
\newblock ``End-to-end neural speaker diarization with permutation-free
  objectives,''
\newblock {\em arXiv preprint arXiv:1909.05952}, 2019.

\bibitem{schroff2015facenet}
Florian Schroff, Dmitry Kalenichenko, and James Philbin,
\newblock ``Facenet: A unified embedding for face recognition and clustering,''
\newblock in {\em Proceedings of the IEEE conference on computer vision and
  pattern recognition}, 2015, pp. 815--823.

\bibitem{panayotov2015librispeech}
Vassil Panayotov, Guoguo Chen, Daniel Povey, and Sanjeev Khudanpur,
\newblock ``Librispeech: an asr corpus based on public domain audio books,''
\newblock in {\em 2015 IEEE International Conference on Acoustics, Speech and
  Signal Processing (ICASSP)}. IEEE, 2015, pp. 5206--5210.

\bibitem{carletta2007unleashing}
Jean Carletta,
\newblock ``Unleashing the killer corpus: experiences in creating the
  multi-everything ami meeting corpus,''
\newblock {\em Language Resources and Evaluation}, vol. 41, no. 2, pp.
  181--190, 2007.

\bibitem{ravanelli2018speaker}
Mirco Ravanelli and Yoshua Bengio,
\newblock ``Speaker recognition from raw waveform with sincnet,''
\newblock in {\em 2018 IEEE Spoken Language Technology Workshop (SLT)}. IEEE,
  2018, pp. 1021--1028.

\bibitem{yin2018neural}
Ruiqing Yin, Herv{\'e} Bredin, and Claude Barras,
\newblock ``Neural speech turn segmentation and affinity propagation for
  speaker diarization,''
\newblock 2018.

\bibitem{nagrani2017voxceleb}
Arsha Nagrani, Joon~Son Chung, and Andrew Zisserman,
\newblock ``Voxceleb: a large-scale speaker identification dataset,''
\newblock {\em arXiv preprint arXiv:1706.08612}, 2017.

\bibitem{chung2018voxceleb2}
Joon~Son Chung, Arsha Nagrani, and Andrew Zisserman,
\newblock ``Voxceleb2: Deep speaker recognition,''
\newblock {\em arXiv preprint arXiv:1806.05622}, 2018.

\bibitem{musan2015}
David Snyder, Guoguo Chen, and Daniel Povey,
\newblock ``{MUSAN}: {A} {M}usic, {S}peech, and {N}oise {C}orpus,'' 2015,
\newblock arXiv:1510.08484v1.

\bibitem{Bredin2020}
Herv{\'e} {Bredin}, Ruiqing {Yin}, Juan~Manuel {Coria}, Gregory {Gelly}, Pavel
  {Korshunov}, Marvin {Lavechin}, Diego {Fustes}, Hadrien {Titeux}, Wassim
  {Bouaziz}, and Marie-Philippe {Gill},
\newblock ``{pyannote.audio: neural building blocks for speaker diarization},''
\newblock in {\em ICASSP 2020}, 2020.

\end{thebibliography}

\end{document}